%% file: spatial-NCSDE-rev2.tex
\documentclass[a4paper]{article}
\usepackage{array,booktabs}
\usepackage{geometry}
\usepackage{amsmath,amssymb}
\usepackage{graphicx}
\usepackage{color,xcolor}
\usepackage{amsfonts,amssymb}
\usepackage{longtable}
\usepackage{subfigure}
\usepackage{hyperref}
\usepackage[english]{babel}
\usepackage[utf8]{inputenc}
\usepackage{amsmath}
\usepackage{graphicx}
\usepackage{animate}
\usepackage{natbib}
\usepackage{epstopdf}
\usepackage[colorinlistoftodos]{todonotes}
\usepackage{colortbl}
\usepackage{float}
\usepackage{amsfonts,amssymb}
\usepackage{multirow}
\include{mydef}

\usepackage{hyperref,arydshln}
\usepackage{siunitx}
\title{Collective Spectral Density Estimation and Clustering for Spatially-Correlated Data}
\makeatletter
\let\@fnsymbol\@arabic
\makeatother
\author{Tianbo Chen$^1$, Ying Sun\thanks{King Abdullah University of Science and Technology (KAUST), Statistics Program, Thuwal 23955-6900, Saudi Arabia. E-mails: tianbo.chen@kaust.edu.sa; ying.sun@kaust.edu.sa.}\ , Mehdi Maadooliat\thanks{Department of Mathematical and Statistical Sciences, Marquette University, USA.  E-mail:mehdi@mscs.mu.edu.}}
\date{}

\begin{document}
\maketitle
\begin{abstract}
\noindent In this paper, we develop a method for estimating and clustering two-dimensional spectral density functions (2D-SDFs) for spatial data from multiple subregions. We use a common set of adaptive basis functions to explain the similarities among the 2D-SDFs in a low-dimensional space and estimate the basis coefficients by maximizing the Whittle likelihood with two penalties. We apply these penalties to impose the smoothness of the estimated 2D-SDFs and the spatial dependence of the spatially-correlated subregions. The proposed technique provides a score matrix, that is comprised of the estimated coefficients associated with the common set of basis functions representing the 2D-SDFs. {Instead of clustering the estimated SDFs directly, we propose to employ the score matrix for clustering purposes, taking advantage of its low-dimensional property.} In a simulation study, we demonstrate that our proposed method outperforms other competing estimation procedures used for clustering. Finally, to validate the described clustering method, we apply the procedure to soil moisture data from the Mississippi basin to produce homogeneous spatial clusters. We produce animations to dynamically show the estimation procedure, including the estimated 2D-SDFs and the score matrix, which provide an intuitive illustration of the proposed method.
\end{abstract}
\newpage
\section{Introduction}

{In spatial statistics, many applications require the segmentation of a spatial region into subregions based on their similarities. Clustering methods are typically developed to address this need. For example, {\cite{ambroise1997clustering} and \cite{allard2000clustering} presented clustering algorithms for spatial data using the EM algorithm}.  \cite{sheikholeslami2000wavecluster} proposed a spatial clustering approach based on wavelet transformation.
{\cite{guillot2006inference} proposed a Bayesian multivariate spatial model to delineate homogeneous regions on the basis of categorical and quantitative measurements.} \cite{tarabalka2009spectral} proposed a spectral-spatial classification scheme for hyperspectral images, which combines the results of a pixel-wise support vector machine classification and the segmentation map obtained by partitional clustering using majority voting.}

{An important challenge in clustering the spatial regions is to take into account the spatial correlation. \cite{romary2015unsupervised} proposed two clustering algorithms based on adaptations of classical algorithms to multivariate geostatistical data, and the spatial dependence is ensured by a proximity condition imposed for two clusters to merge. \cite{fouedjio2016hierarchical} developed an agglomerative hierarchical clustering approach that takes into account the spatial dependency between observations. \cite{fouedjio2017spectral} introduced a spectral clustering approach to discover spatially contiguous and meaningful clusters in multivariate geostatistical data, in which spatial dependence plays an important role. \cite{marchetti2018spatial} proposed to compress the spatial data using spatial dispersion clustering, which produce contiguous spatial clusters and preserve the spatial-correlation structure of the data so that the loss of predictive information is minimal.}

{Note that most of the existing clustering algorithms aim at clustering spatial observations based on similarity of the mean values. Furthermore, spatial processes from real applications are often second-order nonstationary \citep{fouedjio2017second, schmidt2020flexible}. Therefore more sophisticated methods are needed to identify stationary spatial regions with similar dependence structures, or spatial patterns. We tackle this problem via collective estimation of the spectral density functions (SDFs) that follow-up with a clustering step in the spectral domain.}

Efficiency of the estimators for the SDFs are important, as the quality of the estimated SDF directly affects the clustering results. In one-dimensional (1D) cases (time series), the periodogram is a nonparametric estimation of the SDF, and the undesirable properties of the periodogram, such as roughness or inconsistency, have led to the development of many other estimators of the SDF.  In order to achieve a consistent estimator, one method suggests smoothing the periodogram across frequencies. For example, \cite{shumway2016time} discussed several periodogram smoothing techniques, including moving-average smoothing and tapering, and proved that the smoothed periodogram has a smaller variability than the raw periodogram. \cite{wahba1980automatic}  developed the optimally smoothed spline (OSS) estimator, and the smoothing parameter is selected to minimize the expected integrated mean square error. The span selection is an important issue in periodogram smoothing.  \cite{lee1997simple} used the unbiased risk estimator to produce the span selector, whereby the selector did not require strong conditions on the spectral density function. Likelihood is another common method for estimating the spectral density. \cite{capon1983maximum} used the maximum-likelihood filter to produce the minimum-variance unbiased estimator of the spectral density function. \cite{chow1985sieve} proposed a sieve for the estimation of the spectral density of a Gaussian stationary stochastic process using likelihood. \cite{whittle1953estimation, whittle1954some} developed the now well-established Whittle likelihood for time series analysis, and this likelihood is constructed from the spectrum and periodogram. In \cite{pawitan1994nonparametric}, the spectral density function is estimated by the penalized Whittle likelihood. Besides nonparametric estimation of the spectrum, the autoregressive (AR) spectral approximation is discussed in
\cite{shumway2016time}. \cite{chan1982spectral} and \cite{friedlander1984modified} used the Yule-Walker method to estimate the spectrum.

In two-dimensional (2D) case, the 2D periodogram shares similar features to the 1D periodogram. Some examples of asymptotic theorems have been studied in  \cite{heyde1993smoothed} and \cite{stein1995fixed}, and many spatial SDF estimation methods have been developed. For example, \cite{kim2000spectral} applied tapering (data filter) to spatial data in order to reduce the bias of the periodogram. \cite{fuentes2002spectral} proposed a nonstationary periodogram and some parametric approaches to estimate the spatial spectral density of a nonstationary spatial process. \cite{fuentes2007approximate} proposed estimation methods for large, irregularly spaced spatial datasets using Whittle likelihood approximation. \cite{ebeling2006asmooth} developed an efficient algorithm for adaptive kernel smoothing (AKS) of 2D data with a changeable kernel functional form.

In this paper, to cluster spatial data that share similar spectral features, we extend the methodology of collective spectral density functions estimation as proposed by \cite{maadooliat2018nonparametric} to two-dimensional case, and take the spatial dependence of the subregions into account to produce homogeneous spatial clusters. To begin, we use a framework similar to principal component analysis (PCA) to construct a low-dimensional basis expansion that explains the similar features of the 2D-SDFs. Then, we estimate the coefficients associated with the set of adaptive basis by maximizing the Whittle likelihood approximation with two penalties: one to control the smoothness of the adaptive basis functions; the other to consider the spatial dependence of the spatially-correlated subregions to provide more homogeneous spatial clusters. We call the estimated coefficients of the basis expansion as score matrix.  Finally, instead of using the estimated 2D-SDFs for clustering, we propose to cluster the spatial data (2D-SDFs) based on the score matrix, which contains sufficient information on the 2D-SDFs but lives in a lower dimension.  

The remainder of the paper is organized as follows. The proposed method for the 2D-SDFs estimation is introduced in Section 2, and the clustering algorithm is presented in Section 3. In Section 4, we present two simulation studies which consider two cases: with and without spatial dependence. In Section 5, we present the analysis of soil moisture data from the Mississippi basin and in Section 6, we summarize the paper.

\section{Methodology}\label{c2}
\subsection{Spectral Density and Periodogram}
In one-dimensional case, let $x_t, t=1, \cdots, l$ denote a zero-mean weakly stationary time series, and let $\gamma(h)$, denote its autocovariance function (ACF) that satisfies $\sum_{h=-\infty}^{\infty}|\gamma(h)|<\infty,$
then $\gamma(h)$ has the following representation
$$\gamma (h)=\int_{-1/2}^{1/2}e^{2\pi i \omega h}f(\omega)d\omega, \text{  } \text{  } \text{  } h=0,\pm1,\pm2...,$$
where $f(\omega)$ is the spectral density function (SDF) of  $x_t$
$$ f(\omega)=\sum_{h=-\infty}^{\infty}\gamma(h)e^{-2 \pi i \omega h}, \text{  } \text{  } -1/2\leq \omega \leq 1/2.$$
The periodogram is a nonparametric estimate of the SDF. For a given time series $x_t$, the periodogram is calculated by $I(\omega_j)=|d(\omega_j)|^2,$ where $d(\omega_j)$ is the discrete Fourier transform (DFT)
$$d(\omega_j)=l^{-1/2}\sum_{t=1}^lx_te^{-2\pi i \omega_j t},\text{  }\text{  }j=0,1,...,l-1,$$
and the frequencies $ \omega_j= j/l$ are called the Fourier or fundamental frequencies.

In the two-dimensional (2D) case, for a stationary spatial process $z({\bf s})$, ${\bf s}\in \mathbb{R}^2$ with ACF $C(\bs)={\rm Cov}\{z(\bx),z(\bx+\bs)\}$, the 2D-SDF is defined as 
$$f(\bomega)=\int_{\mathbb{R}^2}{\rm exp}(-2\pi i \bomega^{\top}\bs)C(\bs)d\bs,$$
where $\bomega=(u,v)^{\top} \in [-1/2,1/2]\times[-1/2,1/2]$.

Suppose that the spatial process is observed on a regular a $n_{1}\times n_{2}$ lattice $D=\{1,...,n_{1}\} \times \{1,...,n_{2}\}$, the 2D periodogram, $I_{n}$, $n=n_{1}n_{2}$, is defined as 
$$I_{n}(\bomega_{j})=\frac{1}{n}\left|\sum_{\bs \in D}z(\bs){\rm exp}(-2\pi i \omega_{j}^{\top}\bs) \right|^{2},\text{ }\text{ }j=0,...,n-1,$$
where $\bomega_{j}=(u_{j_{1}},v_{j_{2}})^{\top}, j_{1} \in \{0,...,n_{1}-1\}, j_{2} \in \{0,...,n_{2}-1\}$.

\subsection{Collective Estimation}
We consider $m$ subregions that are located on a regular rectangular lattice. Let $z_i(\bs)$, $i=1,..,m$ be the observations in the $i$-th subregion and $f_i$ be the associated 2D-SDF, where $\bs=(x,y)$, $x=y=1,...,n_{1}$ and the size of the subregion is $n=n_1^2$. We propose to estimate the spectral density functions collectively using two sets of basis functions.

We assume that the 2D log-SDFs can be represented by a linear combination of a set of linear independent common basis functions $\{\phi_k(\boldsymbol{\omega}), k=1, \cdots, K\}$ due to the similar features they share. Specifically,
\beq\label{eq:adapt-basis2x} 
f_i( \boldsymbol{\omega}) = \exp\{u_i(\boldsymbol{\omega})\} = \exp \Biggl\{\sum_{k=1}^K \phi_k(\boldsymbol{\omega})\alpha_{ik}\Biggr\}, \qquad i =1, \dots, m, 
\eeq 
where  $\alpha_{ik}$ is the score. The value of $K$ should be a small number so that the number of coefficients can be on a reasonable scale even if $m$ is large.

The common basis functions are not prespecified and need to be determined from the data. We suppose that these common basis functions are constructed using linear combination of a rich family of basis functions, $\{b_\ell(\boldsymbol{\omega}), \ell=1, \cdots, L\}$ ($L \gg K$), such that 
\beq\label{eq:fixed-basisx} 
\phi_k( \boldsymbol{\omega}) = \sum_{\ell=1}^L b_\ell( \boldsymbol{\omega}) \theta_{{\ell}k}, \qquad k = 1\dots, K. 
\eeq 
A large $L$ ensures that the rich basis functions can represent the 2D-SDFs flexibly.

We denote the basis functions and their coefficients:  $\bphi(\boldsymbol{\omega}) = \{\phi_1( \boldsymbol{\omega}), \cdots, \phi_K(\boldsymbol{\omega}) \}^\top$, $\balpha_i=(\alpha_{i1}, \cdots, \alpha_{iK})^\top$, $\bb( \boldsymbol{\omega}) = \{b_1( \boldsymbol{\omega}), \cdots, b_L( \boldsymbol{\omega})\}^\top$, and $\btheta_k =(\theta_{1k}, \cdots, \theta_{Lk})^\top$. We rewrite \eqref{eq:adapt-basis2x} and \eqref{eq:fixed-basisx} into the matrix form $\bU=\bB\bTheta\bA^\top,$ where $\bU=\{u_{1}(\bomega),\cdots,u_{m}(\bomega)\}$ is an $n \times m$ matrix that represents the 2D log-SDFs,  $\bTheta = (\btheta_1, \cdots, \btheta_K)$, and the score matrix $\bA = (\balpha_1, \dots, \balpha_m)^\top$. $\bB=\{\bb(\bomega_1),\cdots,\bb(\bomega_n)\}^\top$ is an $n\times L$ matrix that represents the rich basis functions. The choice of $\bB$ is flexible. In this paper, $\bB$ is the 2D B-spline basis functions matrix which is introduced in Section \ref{penaltyz}. We denote the unknown parameters by $(\bTheta, \bA)$.

\subsection{Whittle Likelihood Approach with Spatial Dependence}
We propose to use the penalized Whittle likelihood that applies the roughness penalty \citep{green1993nonparametric} and spatial dependence penalty to estimate the unknown parameters $(\bTheta, \bA)$:
\begin{equation}\label{eq:pen-likx}
- 2\, \ell_W(\bTheta, \bA) + \lambda_1 {\sf PEN_1}(\bphi)+\lambda_2{\sf PEN_2(\bA)},
\eeq
where
$$\ell_W(\bTheta, \bA) = \sum_{i=1}^m\sum_{j=1}^n\left[u_i(\bomega_{j}) + I_{i,n}(\bomega_{j})\exp\{-u_i(\bomega_{j})\}\right]$$
is the Whittle likelihood approximation  \citep{Whittle1954} and $I_{i,n}$ is the 2D periodogram for the $i$-th subregion. The basis roughness penalty ${\sf PEN_1}(\bphi)$ is used to regularize the basis function to ensure that $\phi_k$  is smooth. Specifically,
\beq \label{penaltyx}
{\sf PEN_{1}}(\bphi) = \sum_{k=1}^K \btheta_k^\top \bR \btheta_k = \tr\{ \bTheta^\top \bR \bTheta\},
\eeq
where the penalty matrix $\bR$ is introduced in Section \ref{penaltyz}.

We consider the spatial dependence of the spatially-correlated subregions using penalty ${\sf PEN_2}(\bA)$. For the $i$-th subregion, we penalize the difference between the basis coefficients of the $i$-th subregion and the nearest subregions.  \cite{sun2016fused} applied a similar approach of penalizing the difference of the estimators based on the spatial locations.  Let $N_i$ be the set of the nearest neighbors of the $i$-th subregion, with $j\in N_i$ representing  the $j$-th subregion as one of the nearest neighbor, excluding the $i=j$ case. Then,
$${\sf PEN_2}(\bA)=\sum_{i=1}^{m}\left| \balpha_i-\frac{1}{|N_i|}\sum_{j \in N_{i}}\balpha_j \right|^2=\sum_{i=1}^{m}\bD_i^{\top}\bD_i,$$
where $\bD_{i }=\balpha_i-\frac{1}{|N_i|}\sum_{j \in N_i}\balpha_j$ and $|N_i|$ is the size of $N_i$, where $|N_i|=2$ if the $i$-th subregion is at corners, $|N_i|$=3 if the $i$-th subregion is on the boundary, and $|N_i|=4$ if otherwise.

The penalized Whittle likelihood approximation is minimized by the Newton-Raphson algorithm. In each iteration, we update $\balpha_i$ for $i = 1, \dots, m$, and $\btheta_k$ for $k=1, \dots, K$ until the convergence.
Specifically,
\beqa\label{eq:update-alphax} 
\balpha_i^{new} & = & \balpha_i^{old} - \tau \, \biggl[\frac{\partial^2\ell_W(\bTheta,\bA)}{\partial \balpha_i \partial \balpha_i^\top} -\lambda_2\frac{\partial^2{\sf PEN_2(\bA)}}{\partial\balpha_i\partial \balpha_i^{\top}}   \biggr]^{-1} \biggl[\frac{\partial\ell_W(\bTheta, \bA)}{\partial \balpha_i} -\lambda_2\frac{\partial{\sf PEN_2}(\bA)}{\partial \balpha_i}\biggr] \bigg\vert_{\bTheta=\bTheta^{old}}^{ \bA=\bA^{old}} \nonumber \\
& = & \balpha_i^{old} - \tau\biggl[\bTheta^\top\sum_{j}\Bigl\{ \bb(\omega_{j}) I_{i,n}(\omega_{j})\exp\bigl[-u_i(\omega_{j})\bigr] \bb(\omega_{j})^\top \Bigr\}\bTheta-\lambda_2\sum_{s=1}^{m}\frac{\partial ^{2}\bD_s^{\top}\bD_{s}}{\partial \balpha_i \partial \balpha_{i}^{\top}}  \biggr]^{-1} \times\nonumber\\
&& \biggl[\bTheta^\top\sum_{j}\Bigl\{ \bb(\omega_{j}) - \bb(\omega_{j}) I_{i,n}(\omega_{j})\exp\bigl[-u_i(\omega_{j})\bigr] \Bigr\}-\lambda_2 \sum_{s=1}^{m}\frac{\partial \bD_s^{\top}\bD_{s}}{\partial \balpha_i}\biggr]\bigg\vert_{\bTheta=\bTheta^{old}, \bA=\bA^{old}} 
\eeqa
and 
\beqa\label{eq:update-thetax} 
\btheta_k^{new} & = & \btheta_k^{old} - \tau \, \biggl[\frac{\partial^2}{\partial \btheta_k \partial \btheta_k^\top} \{\ell_W(\bTheta, \bA)  \} - \lambda_1 \bR \biggr]^{-1}\biggl[\frac{\partial}{\partial \btheta_k} \{ \ell_W(\bTheta, \bA) \} - \lambda_1 \bR \btheta_k\biggr]\bigg\vert_{\bTheta=\bTheta^{old}, \bA=\bA^{old}}\nonumber \\
& = & \btheta_k^{old} - \tau\biggl[\sum_{i=1}^m\alpha_{ik}^2\sum_{j}\Bigl\{ \bb(\omega_{j}) I_{i,n}(\omega_{j})\exp\bigl[-u_i(\omega_{j})\bigr] \bb(\omega_{j})^\top \Bigr\} - \lambda_1 \bR\biggr]^{-1} \times\nonumber\\
&& \qquad \biggl[\sum_{i=1}^m\alpha_{ik}\sum_{j}\Bigl\{ \bb(\omega_{j}) - \bb(\omega_{j}) I_{i,n}(\omega_{j})\exp\bigl[-u_i(\omega_{j})\bigr] \Bigr\} - \lambda_1 \bR \btheta_k\biggr]\bigg\vert_{\bTheta=\bTheta^{old} \bA=\bA^{old}} 
\eeqa 
where the learning rate $\tau$ is the first element in the sequence $\{(1/2)^\delta, \delta=0, 1, \dots\}$, which reduces the penalized Whittle likelihood approximation. We denote the estimator of ($\bTheta$, $\bA$) by ($\widehat{\bTheta}$, $\widehat{\bA})$.

If we only focus on the spectral properties of the subregions where the spatial dependence is not considered, we use
\begin{equation}\label{nospa1}
- 2\, \ell_W(\bTheta, \bA) + \lambda_1 {\sf PEN_1}(\bphi)
\end{equation}
instead of \eqref{eq:pen-likx}, which is same as setting $\lambda_2=0$ in \eqref{eq:update-alphax}. We denote the estimated coefficients from \eqref{nospa1} as $\tilde{\bTheta}$ and $\tilde{\bA}$. The comparison of the clustering results using $\widehat{\bA}$ and $\tilde{\bA}$ is given in Sections \ref{c3} and \ref{c4}.

\subsection{Selecting the Tuning Parameters}
We select $\lambda_1$ and $\lambda_2$ by minimizing the Akaike information criterion (AIC) introduced by \cite{Akaike1974},
$${\rm AIC}(\lambda_1,\lambda_{2}) = - 2\, \ell_W(\widehat{\bTheta}, \widehat{\bA}) + 2\{{\sf df}(\lambda_1) +  {\sf df}(\lambda_2)\}.$$
The degrees of freedom ${\sf df}(\lambda_1)$ and ${\sf df}(\lambda_2)$ are defined as
$$ {\sf df}(\lambda_1)=\sum_{k=1}^K\trace\Bigl\{\bigl[ \frac{\partial^2}{\partial \btheta_k\partial \btheta_k^{\top}}\{\ell_W(\bTheta, \bA)\}-\lambda_1 \bR  \bigr]^{-1}\bigl[ \frac{\partial^2}{\partial \btheta_k\partial \btheta_k^{\top}}\{\ell_W(\bTheta, \bA)\} \bigr]   \Bigr\},$$
and
$$ {\sf df}(\lambda_2)=\sum_{i=1}^m\trace\Bigl\{\bigl[ \frac{\partial^2}{\partial \balpha_i\partial \balpha_i^{\top}}\{\ell_W(\bTheta, \bA)\}-\lambda_2\frac{\partial^2{\sf PEN_2(\bA)}}{\partial\balpha_i\partial \balpha_i^{\top}}\bigr]^{-1}\bigl[ \frac{\partial^2}{\partial \balpha_i\partial \balpha_i^{\top}}\{\ell_W(\bTheta, \bA)\} \bigr]   \Bigr\},$$
in which the parameters are replaced by the estimated values.

Since that it is computationally expensive to search the optimal $\lambda_1$ and $\lambda_2$ by training the model multiple times on sequences of $\lambda_1'$s and $\lambda_2'$s, we update them within the Newton-Raphson
iterations. This method has been described by \cite{Schall1991}, \cite{Schellhase2012}, and \cite{Najibi2017}, where in $p$-th iteration we update
$${\lambda_1}^{(p+1)} =\frac{{\sf df}\{{\lambda_1}^{(p)}\}-(a-1)}{\trace \{\hat\bTheta^{(p)\top} \bR \hat\bTheta^{(p)}\}},$$
and
$${\lambda_2}^{(p+1)}=\frac{{\sf df}({\lambda_2}^{(p)})}{\sum_{i=1}^{m}\left| \balpha^{(p)}_i-\frac{1}{|N_i|}\sum_{j \in N_i}\balpha_j^{(p)} \right|^2} ,$$
where $a=2$ provides the second-order difference penalty given in Section \ref{penaltyz}.

\subsection{2D Basis and Penalties}\label{penaltyz}
We choose  2D spline basis functions as $\bB$ in this paper. Suppose that $\bB^*_l$ is the marginal 1D B-spline basis matrix with $l$ basis  functions of order 4 (to ensure piecewise cubic), then, in \eqref{eq:adapt-basis2x}, $\bB=\bB^*_l\otimes \bB^*_l$, where the number basis functions of $\bB$ is $L=l^2$ and $\otimes$ is the Kronecker product.

We use the spatial roughness penalty matrix $\bR$ to control the roughness of common basis $\phi_k$ using the second-order difference penalty \citep{eilers1996} to achieve the appropriate level of smoothness. The marginal penalty matrix ${\bf r}_l=\bL_l^\top\bL_l$, where
$$ \bL_l= {\left[ {\begin{array}{*{24}{c}} 1&{ - 2}&1& 0& \ldots &0\\    0 & 1 &{ -2}&1 & \ddots &0\\    \vdots & \ddots &\ddots& \ddots & \ddots &0\\    0& \cdots& 0 &1& {-2}&1 \end{array}} \right]_{(l-2) \times l}}.$$ 
Then, the roughness penalty matrix $\bR$ in \eqref{penaltyx} and \eqref{eq:update-thetax} has the representation:
$${\bf R}= {\bf I}_l\otimes {\bf r}_l + {\bf r}_l \otimes  {\bf I}_l,$$
where ${\bf I}_l$ is the identity matrix.

\section{Clustering Algorithm}\label{c2}
We propose to cluster spatial regions based on the estimated score matrix $\widehat{\bA}$, which has the following advantages. First, $\widehat{\bA}$ significantly reduces the dimension from $m\times n$, which is the dimension of the $m$ 2D-SDFs, to $m\times K$. Then, by using singular value decomposition (SVD), we obtain the common basis functions from the rich basis functions, and the property of SVD ensures $\widehat{\bA}$ contains sufficient information. Finally, by considering the spatial dependence using ${\sf PEN_2}(A)$ in \eqref{eq:pen-likx}, we obtain more homogeneous spatial clusters.

A critical step in clustering real data is to identify the number of clusters, which is directly related to the choice of $K$. We use the elbow method \citep{Thorndike1953}, which is widely used in clustering analysis to choose the number of clusters.  To begin, we obtain the smoothed log-periodogram estimation $\bU_{sp}={\bf B}({\bf B}^{\top}{\bf B})^{-1}{\bf B}^{\top} \log ({\bf I})$. In the elbow method, we run a hierarchical clustering method for the smoothed log-periodograms, and compute the total within-cluster sum of squares (WSS) corresponding to the number of clusters $k$. Then, by plotting WSS$(k)$ against $k$, the optimal number of clusters $K$ is found at the location of the elbow or turning point of the plot (see Figure \ref{simulation11}(a) and (b) for illustration). Alternatively, we can also use the Calinski-Harabasz index \citep{calinski1974dendrite} to identify the number of clusters. The Calinski-Harabasz index $ch(k) = \frac{(m-k){\rm tr}(W_1)}{(k-1){\rm tr}(W_2)}$, where $W_1$ is the covariance matrix between clusters and $W_2$ is the covariance matrix within the clusters. The optimal number of clusters is chosen at $K = {\rm argmax}_k ch(k)$.

Below is the clustering algorithm:
\begin{enumerate}
	\item For the $m$ subregions, we obtain the smoothed log-periodogram matrix $\bU_{sp}$, and use the  elbow method (or the Calinski-Harabasz index) based on $\bU_{sp}$ to obtain the optimal number of clusters $K$.
	\item We apply the proposed estimation method, using $K$ common basis functions and obtain $\widehat{\bA}$.
	\item We measure the importance (weights) of the columns of $\widehat{\bA}$ using the singular values. By denoting $w_k$ as the $k$-th singular value and $\widehat{\ba}_k$ as the $k$-th column of $\widehat{\bA}$ ($k=1,...,K$), we have the weighted score matrix $\widehat{\bA}^*=(\widehat{\ba}_1^*,...,\widehat{\ba}_K^*)$, where
	\begin{equation}\label{weight} \widehat{\ba}_k^*=\frac{w_k}{\sum_{k=1}^K w_k}\widehat{\ba}_k.\end{equation}
	\item We compute the Euclidean distance between rows of the matrix $\widehat{\bA}^{*}$  and apply a hierarchical clustering algorithm to the distance matrix using Ward's measure as an agglomeration  method (function {\sf hclust} in the {\sf R} package {\sf stats}).  Where we did not consider the spatial dependence (see Section \ref{sim1} for example), we use $\tilde{\bA}$ instead of $\widehat{\bA}$, then we obtain the weighted score matrix $\tilde{\bA}^*$, and use $\tilde{\bA}^*$ for clustering. Alternative inputs for clustering include the score matrix (without weights) and the estimated 2D-SDF matrix (see the competitive estimators in Section \ref{sim1}).
\end{enumerate}

\section{Simulation Study}\label{c3}
In this section, we perform two simulation studies:  i) a simple case with a known number of clusters without spatial dependence consideration and the estimations are evaluated by clustering results; ii) the subregions are located on a regular grid and the spatial dependence is considered.

We generate the spatial data from a zero-mean Gaussian process with Mat\'ern covariance function: 
$$C(d;\nu,\rho)=\frac{2^{1-\nu}}{\Gamma(\nu)} \left(\sqrt{2\nu}\frac{d}{\rho} \right)^{\nu}K_{\nu}  \left(\sqrt{2\nu}\frac{d}{\rho} \right),$$
where $d$ is the distance, $\Gamma$ is the gamma function, $K_{\nu}$ is the modified Bessel function, $\rho$ is the scale parameter, and $\nu$ is the smoothness parameter.

\subsection{Subregions with Known Number of Clusters and No Spatial Dependence}\label{sim1} 
In this simulation study, we assume that there are three clusters with the same number of subregions. The scale parameters and the smoothness parameters of the Mat\'ern covariance function that we used to generate the subregions in the three clusters are different. Specifically, we consider eight scenarios constructed by four different number of subregions $m=30, 60$ ({to represent small numbers of subregions}), {and} ${m =\ }480, 960$ ({to mimic large numbers of subregions}); and two parameter settings for the Mat\'ern covariance functions:
\begin{itemize}
  \item $p_1$: in the $i$-th cluster, $\rho_i=0.4\times i$ and $\nu_i=0.4\times i$.
  \item $p_2$: in the $i$-th cluster, $\rho_i=0.4\times i$ and $\nu_i=0.4\times(4-i)$.
\end{itemize}
Figure \ref{simulation11}(a) and (b) illustrate the elbow methods of the two parameter settings when $m=30$, where there are turning points at $K=3$, which is in agreement with our cluster setting. We consider three estimators from the proposed method and three competitive estimators for clustering, where the estimators are treated as input in step 4 of Section \ref{c2}:
\begin{itemize}
  \item The estimators from the proposed method ($\tilde{\bA}^{*}$, $\tilde{\bA}$, and estimated spectral density function matrix  $\mathrm{SDF}={\rm exp}(\bB\tilde{\bTheta}\tilde{\bA}^{\top})$).
  \item Smoothed periodograms using the rich basis functions (SPB).\\
  We use the rich basis functions $\bB$ to smooth ${\rm log}({\bf I})$ and obtain ${\rm SPB}={\rm exp}\{{\bf B}({\bf B}^{\top}{\bf B})^{-1}\bB^{\top} \log({\bf I})\}$ as the first competitive estimator.
  \item Smoothed periodograms using 2D Gaussian kernel smoothing (SPK).\\
  We apply 2D Gaussian Kernel smoothing (the bandwidth is selected by generalized cross-validation) to ${\bf I}$ and obtain the second competitive estimator SPK.
  \item Score matrix of the separate estimations ($\tilde{\bA}_{sep}$).\\
  For the $m$  subregions, we maximize the Whittle likelihood separately to obtain the log-SDFs which is an $n \times m$ matrix. We use the truncated SVD of the log-SDFs to obtain the rank $K$ approximation $\bB\tilde{\bTheta}_{sep}\tilde{\bA}_{sep}^{\top}$. Then,  we have the third competitive estimator $\tilde{\bA}_{sep}$.
\end{itemize}

We first measure the performance of clustering by the adjust Rand index (ARI) introduced in  \cite{nguyen2009information}, which is commonly used to compare two clustering results. Note that the ARI ranges from 0 to 1, with 0 indicating that the two clusters do not agree on any pairs and 1 indicating that the clusters are exactly the same. The definition of ARI is:
$${\rm ARI} = \frac{ \sum_{i=0}^1\sum_{j=0}^1 \binom{n_{ij}}{2} - \Bigl[\sum_{i} \binom{n_{i\cdot}}{2} + \sum_{j} \binom{n_{\cdot j}}{2}\Bigr]  / \binom{m}{2} }{ \frac{1}{2} \Bigl[\sum_{i} \binom{n_{i\cdot}}{2} + \sum_{j} \binom{n_{\cdot j}}{2}\Bigr] - \Bigl[\sum_{i} \binom{n_{i\cdot}}{2} + \sum_{j} \binom{n_{\cdot j}}{2}\Bigr]  / \binom{m}{2} }.$$
To calculate the ARI, we compute the $2\times 2$ table, consisting of the following four cells:
		\begin{itemize}
		\item $n_{11}$: the number of observation pairs where both observations are comembers in both clusterings.
		\item $n_{10}$: the number of observation pairs where the observations are comembers in the one clustering but not in the other.
		\item $n_{01}$: the number of observation pairs where the observations are comembers in the second clustering but not in the other.
		\item $n_{00}$: the number of observation pairs where neither pair are comembers in either clustering results.
		\end{itemize} 
We also use the Jaccard coefficients \citep{jaccard1912distribution}{, which is available} in the {\sf R} package {\sf clusteval}, to {further evaluate} the clustering results.
				
In each simulation run, we generate the subregions for each scenario, and obtain the estimators using the proposed method {($\tilde{\bA}^*$, $\tilde{\bA}$, SDF)} and the three competitive estimators {(SPB, SPK, $\tilde{\bA}_{sep}$)}. The clustering results {of} the eight scenarios and six {different} estimators are compared via the true clusters using ARIs and Jaccard coefficients. The {associated} results {(mean ARIs and Jaccard coefficients based on $N=100$ simulation runs)} are {given} in Table \ref{tab1}, in which we can see that the estimators from the proposed method (especially $\tilde{\bA}^{*}$) clearly outperform the {other} competitive estimators in the clustering task. Also, the values {of the clustering indexes (ARIs and Jaccard coefficients) associated to the} scenarios $p_{1}$ are higher {in comparing to the} scenarios $p_{2}$, which is reasonable since the turning point, as shown for two randomly selected simulation runs, in Figure  \ref{simulation11} (a) is much  clearer and sharper than that in Figure \ref{simulation11} (b). Additionally, {as $m$ (the number of subregions) is increasing, the clustering indexes also get closer to one}. We randomly pick a subregion in each cluster associated to the scenario $p_1$, $m=30$ in the first simulation run and use animations to show how the algorithm update the log-SDFs in Animation 1 of the supplementary file. We observe that the power in the low-frequency area (middle) is more dominant when the scale and smoothness parameters increase, which matches the patterns in the corresponding subregions that are shown in Figure \ref{simulation11} (c)-(e) for a randomly selected simulation run.

\begin{table}[!ht]
\begin{center}
\def\arraystretch{0.4}
\begin{tabular}{cccccccc}
\hline
  Scenario   & Measure&   $\tilde{\bA}^*$   &$\tilde{\bA}$& SDF &   SPB &  SPK &  $\tilde{\bA}_{sep}$  \\\hline
 $p_1$, $m=30$  &   ARI   & {\bf 1.000}  & {\bf 1.000}  & {\bf 1.000}  &  0.9844 & 0.9695 & 0.9923 \\ 
    {(3.69 s)} &   Jaccard   & {\bf 1.000}  & {\bf 1.000}  & {\bf 1.000}  &  0.9808 & 0.9619 & 0.9907 \\ \hdashline
 $p_1$, $m=60$ &   ARI   & {\bf 1.000}& {\bf 1.000}   & {\bf 1.000}     &   0.9927 & 0.9742  & 0.9960 \\
    {(6.82 s)} &   Jaccard   & {\bf 1.000}  & {\bf 1.000}  & {\bf 1.000}  &  0.9908 & 0.9678 & 0.9949 \\ \hdashline
 $p_1$, $m=480$  &   ARI  & {\bf 1.000}& {\bf 1.000}   & {\bf 1.000}     &   0.9999 & 0.9915  & 0.9998  \\ 
    {(51.69 s)} &   Jaccard   & {\bf 1.000}  & {\bf 1.000}  & {\bf 1.000}  &  0.9999 & 0.9888 & 0.9998 \\ \hdashline
 $p_1$, $m=960$  &   ARI  & {\bf 1.000}& {\bf 1.000}   & {\bf 1.000}     &   {\bf 1.000} & 0.9891  & 0.9994  \\ 
    {(105.70 s)} &   Jaccard   & {\bf 1.000}  &  {\bf 1.000}  & {\bf 1.000}  &  {\bf 1.000} & 0.9857 & 0.9992 \\\hline
 $p_2$, $m=30$  &   ARI  & {\bf 0.9431}  & 0.8779  &0.8943 & 0.5483 & 0.5018 & 0.3698  \\
    {(3.65 s)} &   Jaccard   & {\bf 0.9304}  & 0.8627  & 0.8753  &  0.5390 & 0.5030 & 0.4201 \\ \hdashline
 $p_2$, $m=60$  &   ARI  &{\bf 0.9465 }  &0.9265    &0.9132  & 0.5410 & 0.4974 & 0.4125   \\
    {(6.90 s)} &   Jaccard   & {\bf 0.9331}  & 0.9114  & 0.8935  &  0.5345 & 0.5032 & 0.4480 \\ \hdashline
 $p_2$, $m=480$ &   ARI   &{\bf 0.9731}   & 0.9676   &0.9037  & 0.5575 & 0.5029 & 0.4518   \\
    {(50.65 s)} &   Jaccard   & {\bf 0.9650}  & 0.9585  & 0.8845  &  0.5497 & 0.5102 & 0.4749 \\ \hdashline
 $p_2$, $m=960$  &   ARI  &{\bf 0.9688}   &0.9667    &0.9170  & 0.5721 & 0.5033 & 0.4763\\
    {(104.39 s)} &   Jaccard   & {\bf 0.9608}  & 0.9582  & 0.8993  &  0.5607 & 0.5114 & 0.4888 \\ \hline
\end{tabular}\\
\end{center}
\label{ri} \linespread{0.5} 
\caption{The clustering results {for $\tilde{\bA}^*, \tilde{\bA}$, SDF, SPB, SPK, and $\tilde{\bA}_{sep}$ using two measures of performance} (ARIs and Jaccard coefficients). The results are based on 100 simulation runs and, in each simulation setup, the best performance is shown in bold. The values within parenthesis, in the first column, provide the average computational time to obtain the collective spectral densities, using a personal computer with $2.6$ GHz Intel Core i$7-9750H$ and $32$ GB memory, in each simulation setup.}
\label{tab1}
\end{table}

\begin{figure}[!ht]
\centering
\includegraphics[width=\textwidth]{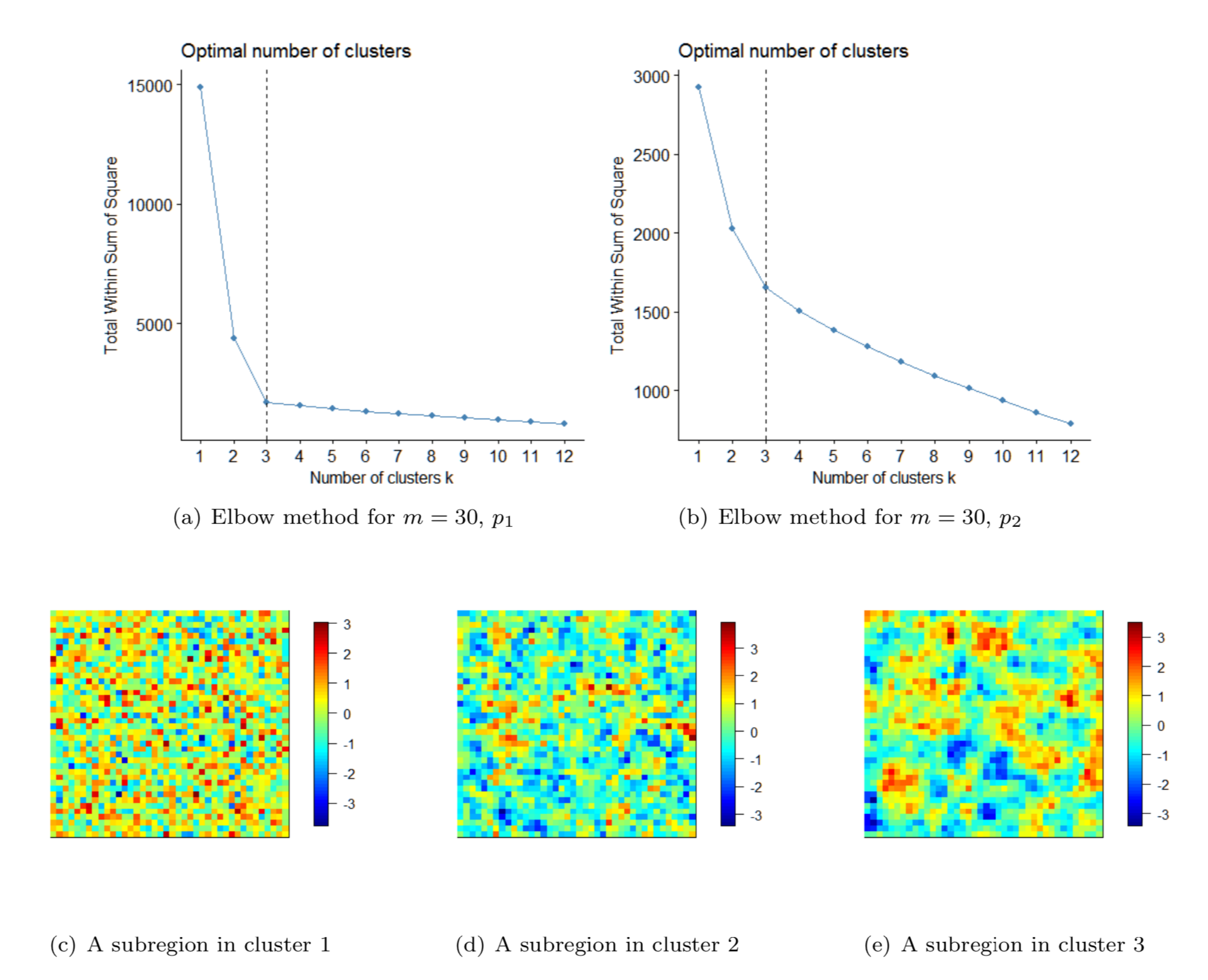}
\linespread{0.5} 
\vspace{-2cm}
\caption{(a) and (b) The elbow method plots for the two parameter settings on two randomly selected simulation runs;  (c) - (e) the  generated subregions ($p_1$, $m=30$) for a randomly selected simulation run.} \label{simulation11}
\end{figure}

\subsection{Clustering with Spatial Dependence and Unknown Number of Clusters}
In this simulation, we perform a more complex case with an unknown number of clusters and the spatial dependence of the subregions is considered. The spatial region  contains $m=$1000 (20 by 50) subregions with different Mat\'ern covariance functions with parameters $\rho$'s and $\nu$'s gradually increasing with the column index and the size of each subregion is 40$\times$40 ($n=1600$). Specifically, $\rho_{col}=\nu_{col}= 0.5+0.05col$,  where $col$ is the column index. Figure \ref{field1} (a) shows the generated random fields, and the elbow method which indicates $K=4$ is shown in Figure \ref{field1} (b) for a randomly selected simulation run.

We apply our proposed method to the subregions and apply the clustering algorithm based on $\widehat{\bA}^{*}$ given that the weighted score matrix had the best performance as outlined in Section \ref{sim1}.  We also estimate $\tilde{\bA}^*$, which does not consider the spatial dependence, for comparison to show the advantage of using $\widehat{\bA}^*$. Figure \ref{field1} (c) and (d) are the clustering results based on $\widehat{\bA}^*$ and $\tilde{\bA}^*$. Both clustering results agree with the increasing trend in the parameters along the horizontal direction, while the proposed method provides more homogeneous clusters: clearer margins, well-separated clusters, and less isolated subregions. We use an animation to dynamically illustrate how the proposed method updating the first column of the score matrix and the corresponding clustering result in Animation 2 of the supplementary file. 

\begin{figure}[!ht]
	\centering
	\includegraphics[width=\textwidth]{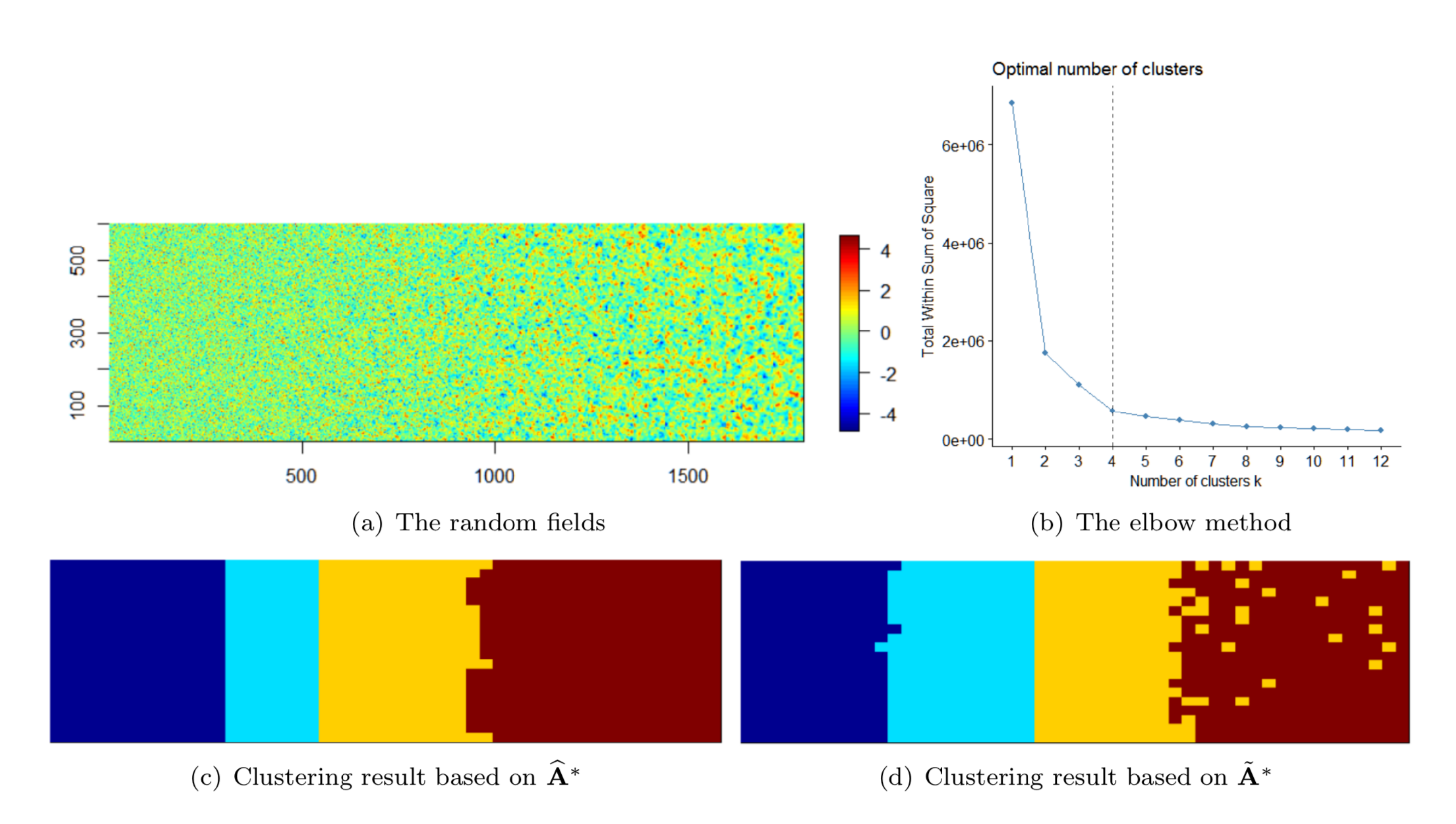}
	\linespread{0.5}
	\vspace{-2.5cm}
		\caption{(a) The random field; (b) the elbow method that indicates $K$=4; (c) and (d) the spatial clustering results based on $\widehat{\bA}^*$ and $\tilde{\bA}^*$ for a randomly selected simulation run.}
	\label{field1}
\end{figure}

\section{Soil Moisture Data Application}\label{c4}
\subsection{Data Description}
Understanding the spatial variability, especially the spatial patterns of  soil moisture is critical for many hydrological applications \citep{brocca2007soil,brocca2012catchment}. In this application, we cluster the soil moisture data of the Mississippi basin area using the proposed method. The location ($92.47^{\circ}-107.72^{\circ}$W, $32.37^{\circ}-43.44^{\circ}$N) of the area is shown in Figure \ref{soilpng} (a) (see \cite{chaney2016hydroblocks} for more details). We consider the soil moisture data for January (winter) and July (summer), and we analyze them separately. For each month, we average 744 (24$\times$31) hourly data and the averaged data are shown in Figure \ref{soilpng} (b) and (c). The size of the region is $1600\times1120$ and we divide the region into $m=1120$ (40 by 28) subregions with size $40\times 40$ ($n=1600$).
\begin{figure}[!ht]
	\centering
	\includegraphics[width=0.8\textwidth]{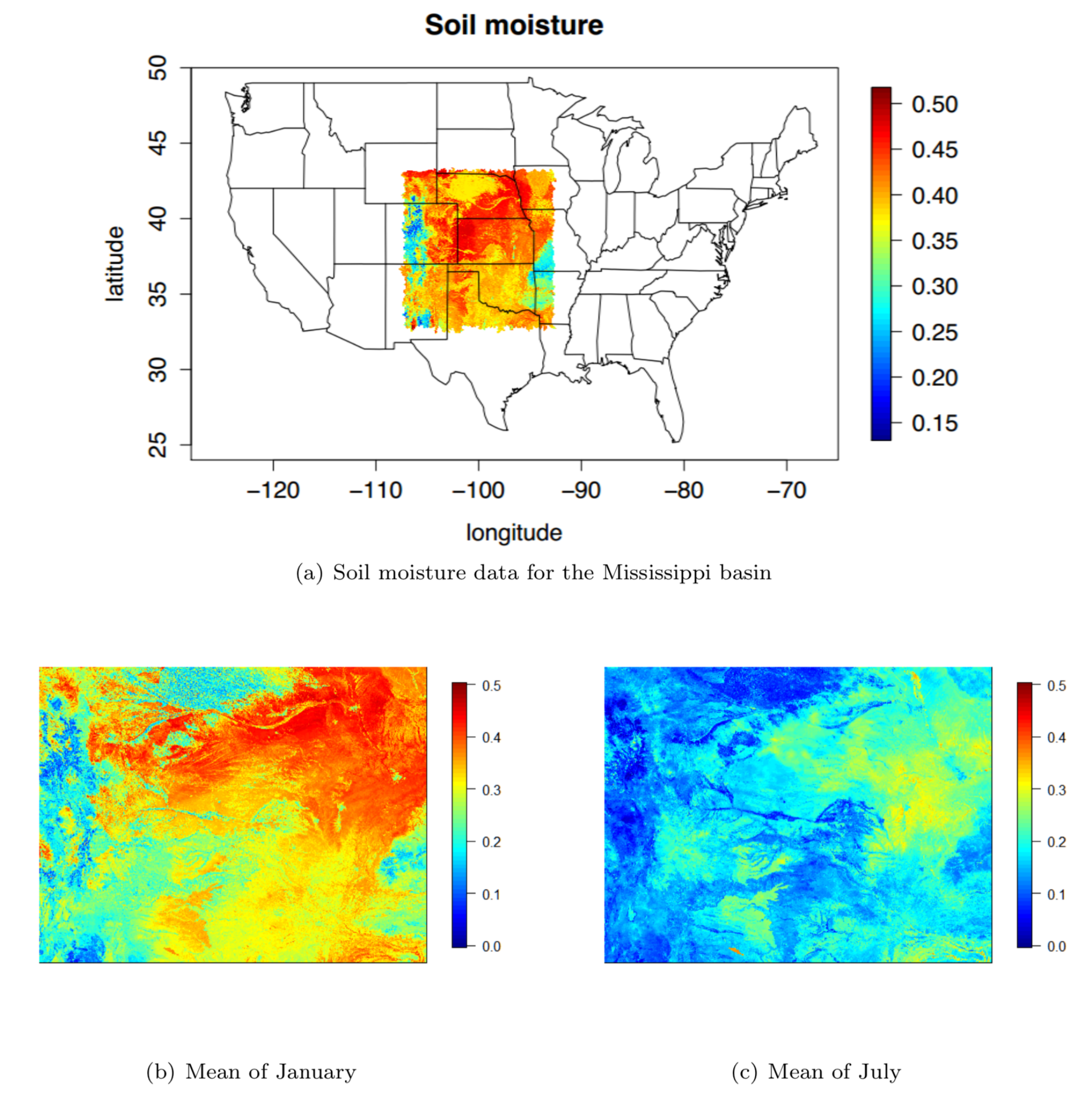}
	\linespread{0.5}
	\caption{(a) The location of the soil moisture data; (b) and (c) the monthly averaged data in January and July (unit: percentage).}
	\label{soilpng}
\end{figure}

\subsection{Clustering Results}
We apply the proposed method to the subregions of the two months, obtaining $\widehat{\bA}^*$, and apply the clustering algorithm. We also do the clustering based on $\tilde{\bA}^*$ for purposes of comparison. {Furthermore, we use the elbow method to identify the number of clusters in the month of January and the Calinski-Harabasz index for the month of July}. Based on the clustering results in Figures \ref{soil1} and \ref{vario}, we obtain the following findings:
\begin{itemize}
\item For the data in January, the elbow method in Figure \ref{soil1} (a) indicates that $K=4$. Out of the 1120 subregions, 361, 134, 437, and 188 subregions are assigned to the four clusters based on $\widehat{\bA}^*$, while 454, 237, 285, and 144 subregions are assigned to the four clusters based on $\tilde{\bA}^*$, respectively. Figure \ref{soil1} (c) and (e) present the corresponding clustering results.
\item For the data in July, the Calinski-Harabasz index in Figure \ref{soil1} (b) indicates that $K=3$. Out of the 1120 subregions, 263, 583, and 274 subregions are assigned to the three clusters based on $\widehat{\bA}^*$, while 342, 531, and 247 subregions are assigned to the three clusters based on $\tilde{\bA}^*$, respectively. Figure \ref{soil1} (d) and (f) present the corresponding clustering results.
\item We observe that the clustering results based on $\widehat{\bA}^*$ have more homogeneous spatial clusters: clearer margins, well-separated clusters, and less isolated subregions, which agree with the animations in Animation 3 of the supplementary file, where the estimation of the score matrices of the two months are illustrated. {However, there are still some spatially non-contiguous subregions. This is due to the fact that clustering results are influenced by the spatial dependence, as well as the similarity of the spectral densities.}
\item For the months of January and July, in Figure \ref{vario}, we present the averaged sample variograms and the associated 95\% confidence intervals of the subregions in each cluster. In Figure \ref{vario} (a), the four clusters are well-separated; while in Figure \ref{vario} (b), the black and red clusters do not have a large difference. We also estimate the parameters of the Mat\'ern covariance function in each subregion using maximum likelihood approach. Then, we applied pairwise two-sample t-test on the estimated coefficients in each of the two clusters. In the case of January, the largest p-value is \num{1.680e-10} and for the month of July, the largest p-value is $0.0227$, which indicates that the coefficients from each of the two clusters are significantly different.
\end{itemize}
\begin{figure}[!ht]
	\centering
	\includegraphics[width=0.8\textwidth]{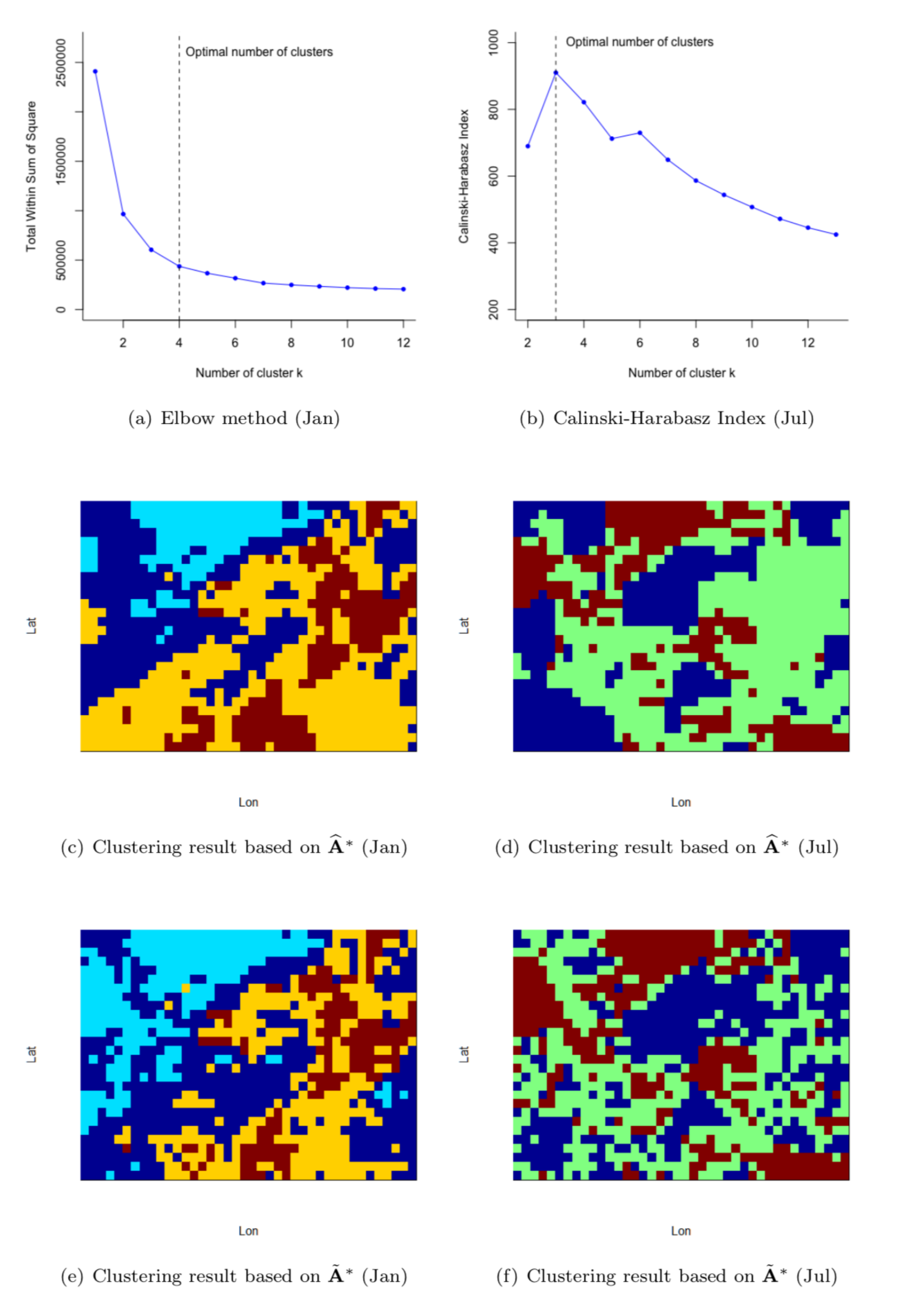}
	\linespread{0.5}
	\vspace{-0.7cm}
	\caption{(a) The elbow method plot (January) and (b) the Calinski-Harabasz Index (July); (c)-(f) the clustering results based on  $\widehat{\bA}^*$ and $\tilde{\bA}^*$ for the two months.}
	\label{soil1}
\end{figure}

\begin{figure}[!ht]
	\centering
	\includegraphics[width=0.95\textwidth]{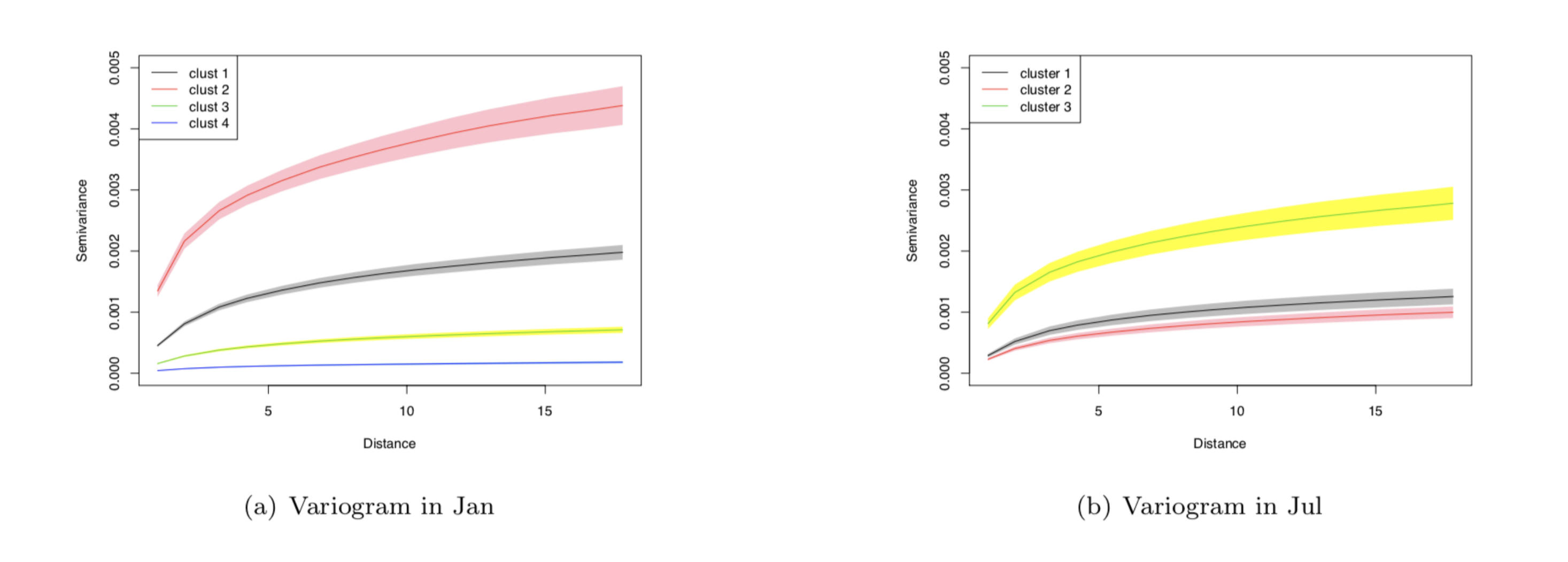}
	\linespread{0.5}
	\vspace{-0.7cm}
	\caption{The averaged sample variograms in each cluster and their 95\% confidence interval; left: January, right: July.}
	\label{vario}
\end{figure}

\section{Conclusion}
In this paper, we developed a highly efficient collective method for 2D-SDFs estimation and clustering. A common set of adaptive basis functions spanned by a rich family of basis was used to explain the similarities among the 2D-SDFs in a lower-dimensional space. The basis coefficients were estimated by maximizing the Whittle likelihood approximation with two penalties using the Newton-type algorithm. One penalty controls the roughness of the basis functions and the other penalty takes the spatial dependence of the spatially-correlated subregions into account. The score matrix, which is the estimated coefficients associated to the basis, is a lower-dimensional representation of the 2D-SDFs which we treated as features to cluster spatial data. The two penalties provide not only smooth estimators of the 2D-SDFs but also more homogeneous spatial clusters. We produce several animations, which intuitively illustrate how the proposed method estimate the 2D-SDFs and the score matrix.

One potential limitation of this paper is that the subregions {are assumed to be} on {a 2D regular} grid. Alternatively one may use more sophisticated 2D-basis, e.g., bivariate splines over triangulations \citep{Maadooliat2016}, that works for complex geometries with unbalanced observations over irregular grid points. Another immediate extension is to introduce the collective estimation approach for multivariate spatial models. 

As for the ease of use, the implementation of the proposed technique is publicly available at \href{https://github.com/tianbochen1/NCSDE_Spatial}{https://github.com/tianbochen1/NCSDE\_Spatial} for reproducing the results of this paper or analyzing any other spatially-correlated dataset.

\section*{Acknowledgment}
The research reported in this publication was supported by funding from King Abdullah University of Science and Technology (KAUST) to Ying Sun and Tianbo Chen. We would also like to thank the editor, and two referees for their constructive and thoughtful comments which helped us tremendously in improving the manuscript.

\bibliography{reff}
\end{document}

%% file: mydef.tex
\def\log{\hbox{log}}

\def\boxit#1{\vbox{\hrule\hbox{\vrule\kern6pt
          \vbox{\kern6pt#1\kern6pt}\kern6pt\vrule}\hrule}}

\def\trace{\hbox{trace}}

\def\bse{\begin{eqnarray*}}
\def\ese{\end{eqnarray*}}
\def\be{\begin{eqnarray}}
\def\ee{\end{eqnarray}}
\def\bq{\begin{equation}}
\def\eq{\end{equation}}
\def\bse{\begin{eqnarray*}}
\def\ese{\end{eqnarray*}}

\newcommand{\bx}{\mathbf{x}}

\newcommand{\bb}{\mathbf{b}}
\newcommand{\bB}{\mathbf{B}}

\newcommand{\bs}{\mathbf{s}}

\newcommand{\balpha}{\boldsymbol{\alpha}}

\newcommand{\btheta}{\boldsymbol{\theta}}
\newcommand{\bphi}{\boldsymbol{\phi}}

\newcommand{\bi}{\begin{itemize}}
\newcommand{\ei}{\end{itemize}}

\newcommand\tr{{\mathrm{tr}}}

\newcommand{\bA}{\mathbf{A}}

\newcommand{\bU}{\mathbf{U}}

\newcommand{\bR}{\mathbf{R}}
\newcommand{\ba}{\mathbf{a}}

\newcommand{\bomega}{\bm{\sigma}}
\newcommand{\bTheta}{\bm{\Theta}}

\def\bx{\mbox{\boldmath $x$}}

\def\bL {\mbox{\boldmath $L$}}

\def\log{\hbox{log}}

\def\boxit#1{\vbox{\hrule\hbox{\vrule\kern6pt
          \vbox{\kern6pt#1\kern6pt}\kern6pt\vrule}\hrule}}

\def\trace{\hbox{trace}}

\def\beq{\begin{equation}}
\def\eeq{\end{equation}}
\def\beqa{\begin{eqnarray}}
\def\eeqa{\end{eqnarray}}
\def\beqs{\begin{equation*}}
\def\eeqs{\end{equation*}}
\def\beqas{\begin{eqnarray*}}
\def\eeqas{\end{eqnarray*}}

\def\boxit#1{\vbox{\hrule\hbox{\vrule\kern6pt
          \vbox{\kern6pt#1\kern6pt}\kern6pt\vrule}\hrule}}
\def\bx{{\ensuremath\mathbf{x}}}

\def\bA{{\bf A}}

\def\bU{{\bf U}}

\def\bx{{\bf x}}

\def\bs{{\bf s}}
\def\ba{{\bf a}}
\def\bb{{\bf b}}

\def\balpha{{\boldsymbol \alpha}}

\def\btheta{{\boldsymbol \theta}}
\def\bomega{{\boldsymbol \omega}}
\def\bphi{{\boldsymbol \phi}}

\def\bTheta{{\boldsymbol \Theta}}

\def\bR{\mathbf{R}}

\def\bB{{\bf B}}
\def\bD{{\bf D}}

\def\bR{{\bf R}}
\def\bU{{\bf U}}

\def\tr{{\mathrm{tr}}}